\documentclass[useAMS,usenatbib]{mn2e}
\usepackage{epsfig}

\def\deg{\ensuremath{^{\circ}}}

\def\<{\ensuremath{<}}
\def\>{\ensuremath{>}}

\title[]{e-VLBI observations of GHz-Peaked Spectrum (GPS) radio sources in nearby galaxies
from the AT20G survey}
\author[Hancock et al.]{
\parbox[t]{\textwidth}
{Paul J. Hancock$^{1}$, Steven J. Tingay$^{2}$, Elaine M. Sadler$^{1}$, Chris Phillips$^{3}$, 
Adam T. Deller$^{4}$}
\vspace*{6pt} \\
$^{1}$Sydney Insitute for Astronomy (SIfA), School of Physics, University of Sydney, NSW 2006, Australia \\
$^{2}$Department of Imaging and Applied Physics, Curtin University of Technology, 
GPO Box U1987, Perth, WA 6102, Australia.\\
$^{3}$Australia Telescope National Facility, CSIRO, P.O. Box 76, Epping, NSW 1710, Australia.\\
$^{4}$Centre for Astrophysics and Supercomputing, Swinburne University of Technology, 
P.O. Box 218, Hawthorn, VIC 3122, Australia.}

\begin{document}
\date{}
\pagerange{\pageref{firstpage}--\pageref{lastpage}} \pubyear{}
\maketitle
\label{firstpage}

\begin{abstract}
GHz-peaked spectrum (GPS) radio sources are thought to be young objects which  
later evolve into FR-I and FR-II radio galaxies.  We have used the Australia Telescope 
20\,GHz (AT20G) survey catalogue to select a uniform sample of GPS sources with spectral 
peaks above 5\,GHz, which should represent the youngest members of this class. 
In this paper, we present e-VLBI observations of ten such objects 
which are associated with nearby ($z<0.15$) galaxies and so represent a 
new population of local, low--power GPS sources.  Our e-VLBI observations were 
carried out at 4.8\,GHz with the Australia Telescope Long Baseline Array (LBA) 
using a real--time software correlator. All ten sources were detected, and were unresolved on scales of $\sim$100\,mas, 
implying that they are typically less than 100\,pc in linear size.  
\end{abstract}

\begin{keywords}
\emph{AGN, GPS, Radio Galaxy Evolution}
\end{keywords}

\section{Introduction}
Gigahertz-Peaked Spectrum (GPS) radio sources are characterized by a spectral peak and turnover 
at frequencies above ~1\,GHz.  They were identified as early as 1966 (Kellerman 1966), and 
are thought to be the progenitors of large radio galaxies (O'Dea 1998).  The spectral 
turnover is usually attributed to synchrotron self--absorption, although free--free absorption 
plays a role in some sources (Tingay \& de Kool 2003, Vermeulen et al. 2003).

Interactions between the central AGN and its host galaxy are especially important in 
the younger sources, as the host ISM plays a large role in the evolution of the radio source. 
In the evolutionary scenario (Snellen et al. 2000), the peak of the radio spectrum progressively 
moves toward lower frequencies as the source evolves.  Most current samples of GPS sources 
are dominated by sources which peak below 5\,GHz, and so are either at large redshift or have 
moved beyond the earliest stages of evolution. 

Recently two very nearby ($d\sim19$\,Mpc) galaxies, NGC\,1052 (Vermeulen et al. 2003) 
and IC\,1459 (Tingay, Edwards \& Tzioumis 2003) have been shown to be GPS radio sources.   
The turnover frequency of the overall spectrum is close to 2.5\,GHz for IC\,1459 and 
10\,GHz for NGC\,1052.  The radio sources within these galaxies are relatively low--powered
($\sim10^{22}$\,W\,Hz$^{-1}$ at 5\,GHz) compared to the ensemble of known GPS sources, 
which have typical radio powers above  $10^{25}$\,W\,Hz$^{-1}$. 
 
Both NGC\,1052 and IC\,1459 (in common with the only other known GPS radio source within 
100\,Mpc, PKS\,1718--649) show strong LINER-like emission lines in their optical spectra 
and have complex gas kinematics in their nuclear regions, which may suggest that the galaxy has 
undergone a recent interaction or gas accretion event.  Franx and Illingworth (1988) note that 
IC\,1459 has a counter-rotating stellar core, which is also postulated to have formed as the 
result of a galaxy merger.  

The GPS radio sources in both NGC\,1052 and IC\,1459 are strongly jet--dominated 
on parsec scales, in contrast to the majority of more distant and luminous GPS radio galaxies 
where the radio emission is dominated by what appear to be the small--scale analogues of 
radio--galaxy hotspots. This raises the possibility that there exists a luminosity/morphology relationship 
in GPS radio galaxies, similar to that seen on much larger scales in FR-I 
and FR-II radio galaxies (Fanaroff \& Riley 1974). 

While the interpretation of NGC\,1052 is unclear, since the radio
source shows the presence of large-scale hotspots indicating that the
radio source is not young, it could be classified as a restarted radio
source.  IC\,1459, however, has no large-scale structure and could be
interpreted as a young radio source.  The apparently two-sided nature
of the pc-scale jets in IC 1459 (Sokolova et al. 2009, in preparation)
does not favour a highly aligned jet as an explanation for the lack of
large-scale structure.

When investigating GPS radio sources it is important to recognize the potential for contamination
from variable sources whose peaked spectrum is not at all related to the evolution of young radio galaxies.
Less than 10\% of the QSOs identified as GPS sources in the literature retain their classification when subjected to 
long term monitoring and simultaneous spectral measurements (Torniainen et al. 2005). Similarly--selected galaxy type
GPS samples are more reliable with $\sim 40$\% identified as genuine GPS sources (Torniainen et al. 2007).
In each case the contamination rate is seen to increase as the spectral peak shifts to higher frequencies.

To investigate the parsec-scale properties of GPS radio sources at the
lowest luminosities further, we have used the high-frequency AT20G survey (Ricci et al. 2004; 
Sadler et al. 2006, Massardi et al. 2008) to construct a uniform sample of GPS sources 
with high--frequency ($>8$\,GHz) spectral turnovers, low redshift ($z<0.15$) and low 
radio power (P$_{5}<10^{24.5}$\,W/Hz). To avoid as much as possible the contamination from 
sources unrelated to the evolutionary scenario, we selected only sources that are identified with galaxies.
The source names refered to in this paper are drawn from the AT20G survey.
In this paper, we present observations from the Australian e-VLBI network at $\sim$100\,mas 
resolution as a first step in measuring the angular sizes and structures of these nearby high-frequency GPS sources.  

We use the following cosmological values throughout this paper: $H_0=71kms^{-1}$, 
$\Omega_m=0.27$, and $\Omega_\Lambda=0.73$.

\section{Target Selection}

\subsection{Selecting high-frequency GPS sources} 
The AT20G full-sample data release (Murphy et al.\ 2009, in preparation) provides 
near-simultaneous flux--density measurements at 4.8, 8.6 and 20\,GHz for most 
AT20G sources south of declination $-15^\circ$.  This is important in allowing us 
to identify candidate GPS sources without the problem of variability giving a false spectral 
shape. 

We selected our high--frequency GPS sample from the $\sim$3800 AT20G sources 
which had good--quality data at all three frequencies, since this gives us enough 
spectral information to identify an inverted or peaked spectrum. All the AT20G sources 
observed at 5 and 8\,GHz were detected at these frequencies, so there are no upper limits 
in the AT20G sample. 
AT20G sources whose radio spectrum peaked above 5\,GHz, or rose with frequency 
over the whole 5--20\,GHz range\footnote{Sadler et al.\ (2008) found that almost all  
AT20G sources with rising (`inverted') spectra at 5--20\,GHz show a spectral turnover 
between 20 and 95\,GHz, so we are confident that most of the ``inverted--spectrum'' 
AT20G sources will be high-frequency peaking GPS objects.}
were flagged as GPS candidates.  This yielded a final list of 656 candidate high--frequency 
GPS sources with spectral peaks above 5\,GHz (Hancock 2009). 
The 1.4\,GHz NVSS (Condon et al.\ 1998) 
and 843\,MHz SUMSS (Mauch et al.\ 2003) catalogues were also used to search for low--frequency 
emission from the AT20G GPS sources. 
%


\subsection{Optical identification}
We cross--matched our AT20G GPS sample with the optical SuperCOSMOS catalogue 
(Hambly et al.\ 2001) to search for optical counterparts.  
Roughly 70\% of our AT20G GPS sample (465/656) had an optical identification above the SuperCOSMOS 
plate limit. Of these 23\% were classified as galaxies by SuperCOSMOS and 77\% were stellar objects, 
which are expected to be QSOs. This is consistent with the finding of Stanghellini (2003) that GPS 
populations contain many flat-spectrum radio QSOs. 

\subsection{A complete sample of nearby GPS radio galaxies} 
Since our interest here is in the GPS radio sources associated with nearby galaxies, we considered 
only the $\sim$100 AT20G GPS sources which were identified with SuperCOSMOS galaxies. A radio--optical
identification was accepted if the two positions differed by less than 5\,arcsec.
Redshifts for these objects were obtained from the 6dF Galaxy Survey (6dFGS; Jones et al. 2004) 
and from the wider literature via the NED online database\footnote{http://nedwww.ipac.caltech.edu/}. 

Only about 25\% of the AT20G GPS galaxies currently have redshift information, 
and further redshift measurements are in progress. 
Using the currently-available redshift data, we identified a sample 
of 28 GPS radio sources which were associated with nearby (redshift $z<0.15$) 
galaxies.  
All of these had 5\,GHz radio powers below $10^{24.5}$\,W\,Hz$^{-1}$.
Ten of the galaxies from this sample (listed in Table \ref{table:observed_targets})
were observed in our March 2008 eVLBI run.  The low--frequency properties of these 
sources are summarized in Table \ref{table:sumss_nvss}. 

\begin{table*}
 \begin{tabular}{crrrrrrrrrrrl}
\hline
 Name & \multicolumn{2}{c}{AT20G position} & ${\rm K_s}$ &  z  & $S_{5}$ & $S_{8}$ & $S_{20}$ & $\alpha_5^8$ & $\alpha_5^{20}$ & log\,P$_5$ & $\nu_{\rm peak}$ & Alt. \\
AT20G & \multicolumn{2}{c}{J2000}          &   mag       &     & mJy     & mJy     & mJy      &		     &		       & WHz$^{-1}$ &   GHz            & name \\
 (1)  &   \multicolumn{2}{c}{(2)}          &   (3)       & (4) & (5)     & (6)     & (7)      &    (8)       &     (9)         &   (10)     &   (11)           & (12) \\
\hline
J031010$-$573041 & 03 10 10.6 & $-$57 30 41.3 & 13.1     & 0.082 &  45  &  73   &   89  & $+$0.82 &	$+$0.48	  & 23.8 &  $\sim$20   & ESO\,116-G10 \\
J051103$-$255450 & 05 11 03.8 & $-$25 54 51.0 & 12.2     & 0.092 &  92  & 113   &  122  & $+$0.35 &	$+$0.20	  & 24.2 &  $\sim$20   & PMN\,J0511$-$2554 \\
J054828$-$331331 & 05 48 28.5 & $-$33 13 31.5 & 12.9     & 0.040 &  31  &  38   &   52  & $+$0.35 &	$+$0.36	  & 23.0 &  \>20       & \dots \\
J074618$-$570258 & 07 46 18.7 & $-$57 02 58.6 & \dots    & 0.130 &  47  &  63   &   94  & $+$0.50 &	$+$0.49	  & 24.2 &  \>20       & SGRS\,J0746$-$5702   \\
J091856$-$243829 & 09 18 56.5 & $-$24 38 29.5 & 14.4     & 0.056 &  48  &  51   &   64  & $+$0.10 &	$+$0.20	  & 23.5 &  \>20       & \dots \\
J114503$-$325824 & 11 45 03.5 & $-$32 58 24.2 & 10.5     & 0.038 &  69  &  91   &   75  & $+$0.47 &	$+$0.06	  & 23.3 &  $\sim$10   & \dots \\
J130031$-$441442 & 13 00 31.1 & $-$44 14 42.6 & 10.5     & 0.032 &  68  & 104   &   77  & $+$0.72 &	$+$0.09	  & 23.2 &  $\sim$10   & AM\,1257$-$435 \\
J181857$-$550815 & 18 18 58.0 & $-$55 08 15.2 & 11.4     & 0.072 &  42  &  53   &   74  & $+$0.41 &	$+$0.40	  & 23.7 &  \>20       & \dots \\
J220916$-$471000 & 22 09 16.3 & $-$47 10 00.3 &  7.1     & 0.005 & 136  & 161   &  122  & $+$0.29 &	$-$0.07	  & 21.8 &  $\sim$10   & NGC\,7213 \\
J224506$-$433157 & 22 45 06.0 & $-$43 31 57.3 & 13.0     & 0.068 &  74  &  90   &   84  & $+$0.33&	$+$0.09	  & 23.9 &  $\sim$10   & \dots\\
\hline							
 \end{tabular}
 \caption{AT20G sources observed in our eVLBI run.  Positions and flux densities are taken from the AT20G Final Data Release 
 (Murphy et al.\ 2009, in preparation). Redshifts are from the Third Data Release (DR3) of the 6dF Galaxy Survey (Jones et al.\ 2009  
 in preparation) except for J074618$-$570258 (Saripalli et al.\ 2005), J130031$-$441442 and J220916$-$471000 (both from de Vaucouleurs et al. 1991). 
Column 3 lists the total K$_{\rm s}$-band magnitude from the 2MASS Extended Source Catalog (Jarrett et al.\ 2000).   
The frequency at which the radio spectrum peaks (column 11) is estimated from the three quasi-simultaneous AT20G data points. 
The errors on the AT20G flux density measurements are typically $\sim$5\%, giving typical uncertainties of $\pm$0.15 
in the radio spectral index $\alpha$. }
 \label{table:observed_targets}
\end{table*}

\begin{table}
 \begin{tabular}{cccl}
\hline 
 Name  & S$_{\rm SUMSS}$         &  S$_{\rm NVSS}$  & log\,P$_{1.4}$  \\
 AT20G & mJy                     & mJy              & WHz$^{-1}$      \\
\hline
J031010$-$573041 &   37.3 $\pm$ 1.5  &  ...             &  23.8     \\  
J051103$-$255450 &   ...             &  44.1 $\pm$  1.7 &  24.0     \\
J054828$-$331331 &  $<$10            &   6.0 $\pm$  0.5 &  22.3     \\
J074618$-$570258 &  194.0 $\pm$ 11.0 &  ...             &  24.8     \\  
J091856$-$243829 &   ...             &  55.6 $\pm$  1.7 &  23.6     \\
J114503$-$325824 &   39.7 $\pm$  2.2 &  31.5 $\pm$  1.1 &  23.0     \\
J130031$-$441442 &   26.9 $\pm$  1.3 &  ...             &  22.9     \\  
J181857$-$550815 &  627.7 $\pm$ 21.2 &  ...             &  24.7     \\  
J220916$-$471000 &  119.6 $\pm$  3.8 &  ...             &  21.8     \\  
J224506$-$433157 &   38.7 $\pm$  1.5 &  ...             &  23.7     \\  
\hline							
 \end{tabular}
 \caption{Low--frequency radio properties of the nearby GPS radio sources in our sample. For galaxies south of 
 declination $-40^\circ$, the 1.4\,GHz radio power is calculated from the 843\,MHz SUMSS flux density 
 and the measured 0.8--5\,GHz spectral index except for the radio galaxies J074618$-$570258 and J181857$-$550815, 
 where we assume a spectral index of $-0.7$ for the extended low-frequency emission. J054828$-$331331 is undetected 
 in the SUMSS catalogue, so only an upper limit is given. }
 \label{table:sumss_nvss}
\end{table}

\section{Observations}
The targets listed in Table \ref{table:observed_targets} were observed using the new e-VLBI capability of the Long Baseline Array 
(LBA, Tzioumis 1997) in March 2008, at a frequency of 4.8GHz. The LBA stations used were the Parkes observatory (64\,m dish), ATCA (six 22\,m dishes 
as a tied array), and the Mopra antenna (22m) of the ATNF. The data were correlated in real time at Parkes using a Distributed FX (DiFX) correlator developed by Deller et. al (2007).

\begin{figure}
 \epsfig{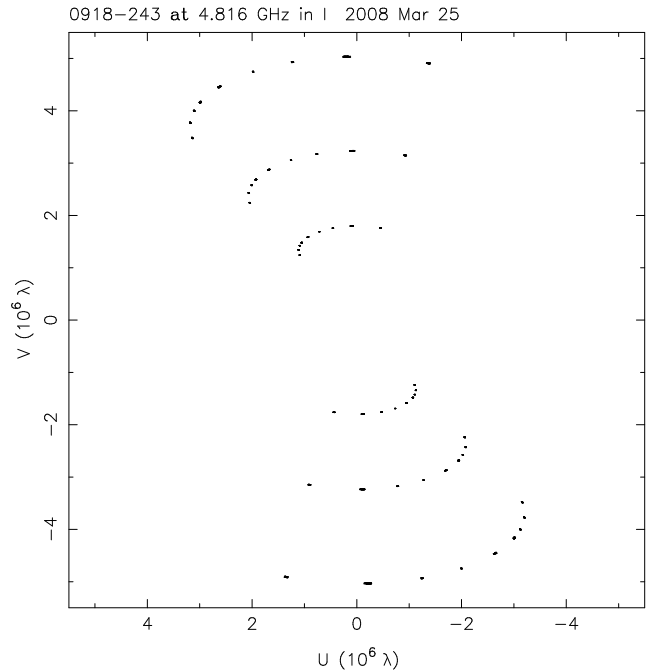}
 \caption{Typical u-v coverage for the e-VLBI observations of the sources within the sample. Each source has approximately 10 $\times$ 5 minute integrations spanning a 12 hour period.}
 \label{fig:uvplot}
\end{figure}

The observations were arranged as a series of 5 minute integrations per source, cycling through those sources that were above the horizon
at any time.  In this way, a typical ensemble of observations for an
individual source consisted of approximately 10 $\times$ 5 minute
integrations, over a 12 hour period.  A typical $uv$ coverage is
shown in figure\,\ref{fig:uvplot}.  These data were reduced using standard VLBI data
reduction and imaging techniques implemented in AIPS\footnote{The
Astronomical Image Processing System (AIPS) was developed and is
maintained by the National Radio Astronomy Observatory, which is
operated by Associated Universities, Inc., under co-operative
agreement with the National Science Foundation} and DIFMAP (Shepherd
1994).  The observations utilised phase-referencing techniques via
observations of bright, compact calibration sources nearby to each
target, for calibration of the interferometer phase, enhancing the
coherence time of the visibilities and allowing the detection of
fainter targets.

The typical one sigma image sensitivity derived from these datasets is
approximately 1 mJy/beam.  The angular resolution varies with
source declination but is typically approximately 100 mas.

\begin{table*}
 \begin{tabular}{crrrrrrrrrrc}
  \hline
 Name   & \multicolumn{2}{c}{eVLBI position} & $\Delta $     & z     & M$_{\rm K}$ & Scale & LAS     & LLS  & S$_{\rm VLBI}$ & S$_{\rm AT20G}$ & Flux ratio  \\
 AT20G  & \multicolumn{2}{c}{J2000}        & arcsec &      & mag & kpc/''     & arcsec  & pc   &  mJy      & mJy        &  \small{ e-VLBI/AT20G} \\
 (1)    & \multicolumn{2}{c}{(2)}  &   (3) &   (4)       &     (5)      &    (6)      &   (7)     &   (8)  &  (9) & (10) & (11)  \\
\hline
J031010$-$573041 & 03 10 10.6 & $-$57 30 41.68 & 0.1 & 0.082 &  $-24.7$ & 1.54  &  \<0.13  & \<200 &  36.5 &  45  &  0.81  \\  
J051103$-$255450 & 05 11 03.8 & $-$25 54 49.96 & 0.5 & 0.092 &  $-25.9$ & 1.71  &  \<0.11  & \<188 &  47.2 &  92  &  0.51   \\ 
J054828$-$331331 & 05 48 28.5 & $-$33 13 30.00 & 1.1 & 0.040 &  $-23.3$ & 0.80  &  \<0.09  & \<72  &  24.5 &  31  &  0.79   \\ 
J074618$-$570258 & 07 46 18.6 & $-$57 02 58.21 & 0.0 & 0.130 &   ...    & 2.29  &  \<0.05  & \<114 &  62.2 &  47  &  1.32   \\ 
J091856$-$243829$^*$ & 09 18 56.5 & $-$24 38 29.39 & 3.7 & 0.056 &  $-22.6$ & 1.07  &  \<0.06  & \<64  &  46.2 &  48  &  0.96   \\ 
J114503$-$325824 & 11 45 03.5 & $-$32 58 23.48 & 0.3 & 0.038 &  $-25.6$ & 0.74  &  \<0.05  & \<37  &  62.9 &  69  &  0.91   \\ 
J130031$-$441442 & 13 00 31.0 & $-$44 14 41.51 & 0.2 & 0.032 &  $-25.2$ & 0.63  &  \<0.06  & \<38  &  58.6 &  68  &  0.86   \\ 
J181857$-$550815 & 18 18 58.0 & $-$55 08 15.30 & 0.3 & 0.072 &  $-26.1$ & 1.35  &  \<0.06  & \<81  &  47.9 &  42  &  1.14   \\ 
J220916$-$471000 & 22 09 16.2 & $-$47 10 00.25 & 0.5 & 0.005 &  $-24.5$ & 0.10  &  \<0.15  & \<15  & 120.6 & 136  &  0.89   \\ 
J224506$-$433157 & 22 45 06.0 & $-$43 31 57.44 & 0.1 & 0.068 &  $-24.4$ & 1.29  &  \<0.20  & \<258 &  56.5 &  74  &  0.76   \\ 
  \hline
 \end{tabular}
 \caption{VLBI core positions and upper limits on angular and linear size for the AT20G sources in Table \ref{table:observed_targets}. 
$\Delta$ (column 3) is the offset between the optical and e-VLBI positions. Columns 9 and 10 list the 4.8\,GHz flux density measured from 
the AT20G ($\sim$15\,arcsec beam) and e-VLBI ($\sim$0.1\,arcsec beam) 
images. $^*$ Note that (as discussed in section \ref{sec:J091856-243829}) this radio source may be a background QSO. }
 \label{table:position_and_size}
\end{table*}

\section{Results}
\begin{figure}
 \epsfig{file=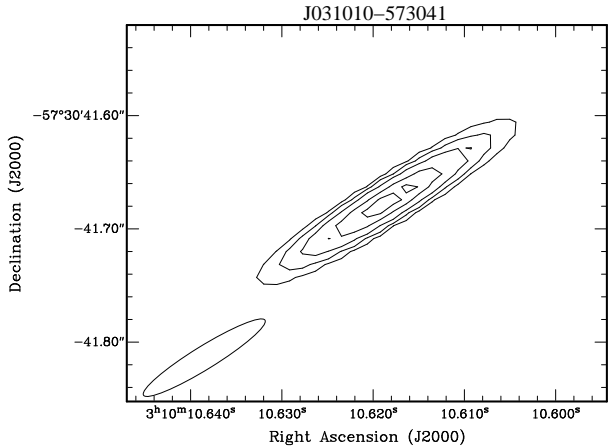, width=8cm}
 \caption{A typical 4.8\,GHz eVLBI image from this program, for the nearby galaxy ESO\,116--G10 
(J031010$-$573041).  The restoring beam is 0.13$\times$0.02\,arcsec in position angle -58\deg, the rms noise 
in the image is 0.7 mJy/beam and the peak flux density is 36.5\,mJy/beam.  
Contour levels are 2, 5, 10, 20 and 30 mJy/beam. The source is unresolved. }
 \label{fig:J031010-573041}
\end{figure}

Figure \ref{fig:J031010-573041} shows a typical e-VLBI image of one of the sources in Table \ref{table:observed_targets}, 
all of which were unresolved on scales of 200\,mas or less. 

Table \ref{table:position_and_size} lists the maximum angular size of each source as measured from the images using the 
MIRIAD task IMFIT.  An e-VLBI position is also listed for each source along with the 4.8\,GHz flux density 
measured from the e-VLBI ($\sim0.1$\,arcsec beam) and AT20G ($\sim15$\,arcsec beam) images. The uncertainty in the 
the e-VLBI positions is dominated by the phase referencing of the calibration source, and is typically 10\,mas in RA and DEC.
Note that the AT20G and e-VLBI flux-density measurements are not simultaneous and were made up to three years apart.

The mean flux ratio (S$_{\rm VLBI}$/S$_{\rm AT20G}$) is 0.90 with a standard deviation of 0.22.  If we exclude the 
source J051103$-$255450, which appears to be variable (see \SS4.1.2 below), the mean flux ratio rises to 0.94 and the
standard deviation drops to 0.18. These results suggest that 
(i) the nearby AT20G GPS sources are compact, with $\sim$90\% of their 4.8\,GHz emission arising on scales smaller 
than 100--200\,pc, and (ii) most of these sources show only modest variability at 4.8\,GHz on timescales of 1--3 years. 

\subsection{Notes on individual sources}

\subsubsection{AT20G J031010$-$573041}
\begin{figure}
 \epsfig{file=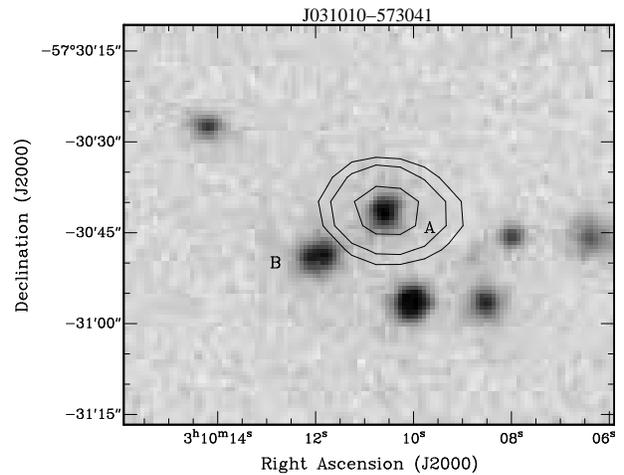, width=8cm}
 \caption{Blue SuperCOSMOS optical image overlaid with AT20G 20\,GHz radio contours at 10, 20 and 50\,mJy/beam. 
As discussed in the text, galaxy A is the AT20G source J031010$-$573041 and galaxy B has a measured redshift of 
$z=0.082$ which is adopted as the redshift of the group.}
 \label{fig:J031010-573041_optical}
\end{figure}

The optical counterpart of this radio source is a member of a compact group of galaxies (Figure \ref{fig:J031010-573041_optical}).  
No redshift has been measured for the host galaxy (object A in 
Figure \ref{fig:J031010-573041_optical}), so we adopt the measured 6dFGS redshift of $z=0.082$ for the companion galaxy 
as the redshift of the whole group.  None of the other galaxies in this group has a redshift measurement.

\subsubsection{AT20G J051103$-$255450}
This source was detected in the PMN survey (Griffith et al. 1994)  with a flux density 
of 53$\pm$11\,mJy in the 4\,arcmin Parkes beam at 4.8GHz.   This is significantly lower 
than the AT20G value of 92$\pm$4\,mJy (in a 15\,arcsec beam), suggesting that the source 
may be variable.  The 6dFGS spectrum shows absorption lines typical of an early--type 
galaxy but no obvious optical emission lines. 

\subsubsection{AT20G J054828$-$331331}
The 6dFGS spectrum shows absorption lines together with possible weak [O\,III] emission. 
There is a faint NVSS source associated with this object (see Table \ref{table:sumss_nvss}), 
but it lies below the limit of the 843\,MHz SUMSS catalogue. 

\subsubsection{AT20G J074618$-$570258}
\begin{figure}
 \epsfig{file=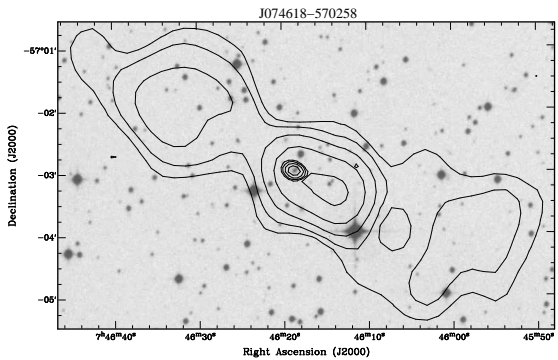, width=8cm} 
 \caption{J074618$-$570258 radio contours overlaid on a blue optical image. SUMSS contours (outer) are at 3, 
 6, 12 and 48 mJy/beam with a beam of 54.3$\times$45\,arcsec. AT20G contours (inner) are at 6, 12 and 48\,mJy/beam 
 with a beam of 12.7$\times$9.2\,arcsec.}
 \label{fig:J074618-570258_sumss}
\end{figure}

This radio source has been identified by Saripalli et al. (2005) as the core of a 
``double-double'' Giant Radio Galaxy. Figure \ref{fig:J074618-570258_sumss} shows the 
optical SuperCOSMOS blue image overlaid with SUMSS radio contours at 843MHz -- 
the total extent of the SUMSS source is 5.5\,arcmin, 
corresponding to a largest linear size of $\sim$750\,kpc. 

This source was classified as FR-I by Saripalli et al. (2005) on the basis of the 
edge--darkened radio morphology in the SUMSS image. 
Saripalli et al.\ (2005) also obtained a higher-resolution 1.4\,GHz radio image of 
J074618$-$570258 with the ATCA (their Figure 9), which shows an inner pair of radio hotspots 
indicative of a core-jet morphology with the jet pointing towards the weaker (South-West) 
of the larger--scale radio lobes. 
The component identified as the core by Saripalli et al. (2005), which is identified with 
a $z=0.13$ galaxy, is also coincident with the AT20G source detected in our eVLBI observation. 

Saripalli et al.\ (2005) note that the optical spectrum of this galaxy (shown in their 
Figure 19) has stellar absorption lines but no obvious emission lines.  They suggest that 
J074618$-$570258 may be an example of a restarting radio jet within relic lobes. 

\subsubsection{AT20G J091856$-$243829}\label{sec:J091856-243829}
The 6dFGS spectrum of this galaxy shows strong emission--lines of H$\alpha$, [N\,II] and 
[S\,II], although H$\beta$/[O\,III] lines are not seen. The e-VLBI position is offset 
3.7\,arcsec north of the optical centroid of the galaxy. 

This galaxy has a K-band absolute magnitude M$_{\rm K}=-22.6$, making it significantly 
less luminous than the host galaxies of most 
nearby radio--loud AGN (which typically have M$_{\rm K}<-24$, see e.g. Figure 
8 of Mauch \& Sadler 2007). 
This, together with the relatively flat radio spectrum (compared to the other 
sources in our GPS sample) and the 3.7\,arcsec radio--optical position offset, 
suggests that the radio emission in J091856$-$243829 may come from 
a background quasar rather than the galaxy itself.  

\subsubsection{AT20G J114503$-$325824}
This galaxy appears in the SUMSS and NVSS catalogues as a point source of 39.7\,mJy and 
31.5\,mJy respectively, suggesting a strong upturn in the radio spectrum above 1.4\,GHz. 
The 6dFGS spectrum shows strong absorption lines and weak [N\,II] emission, consistent with 
the galaxy's designation as a possible LINER. 

\subsubsection{AT20G J130031$-$441442}
\begin{figure}
 \epsfig{file=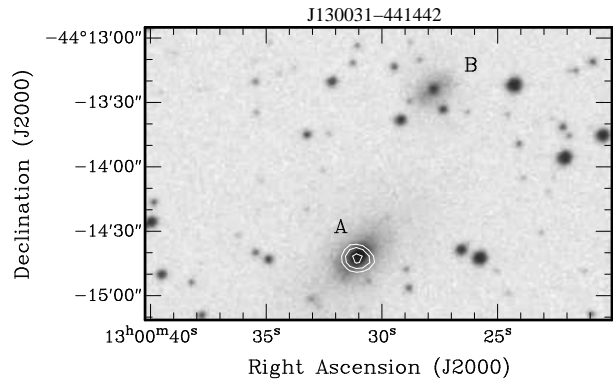, width=8cm}
 \caption{SuperCOSMOS blue image with the galaxy pair AM\,1257$-$435 identified. Galaxy A is the 
 radio source and B the companion galaxy discussed in the text. AT20G 20GHz contours are overlaid at 20, 40, and 80\,mJy/beam. }
 \label{fig:J130031-441442_optical}
\end{figure}

This object is associated with the brighter of the two objects in the galaxy pair AM\,1257$-$435 
(galaxy A in Figure \ref{fig:J130031-441442_optical}). Galaxy B, the second member of the pair, is 
1.4\,arcmin away. 
J130031$-$441442 (galaxy A) is listed as a shell galaxy in the catalogue of Malin and Carter (1983), who 
describe it as having ``shells NW and SE, 2 companions''.  Such shells are generally attributed 
to a past merger of two gas--poor galaxies. 
J130031$-$441442 is also identified with the UV source FC-238 by Brosch et al. (2000), who list it as a 
SAB(s) pec galaxy with UV magnitude of $12.58\pm0.61$.  
 
\subsubsection{AT20G J181857$-$550815}
\begin{figure}
 \epsfig{file=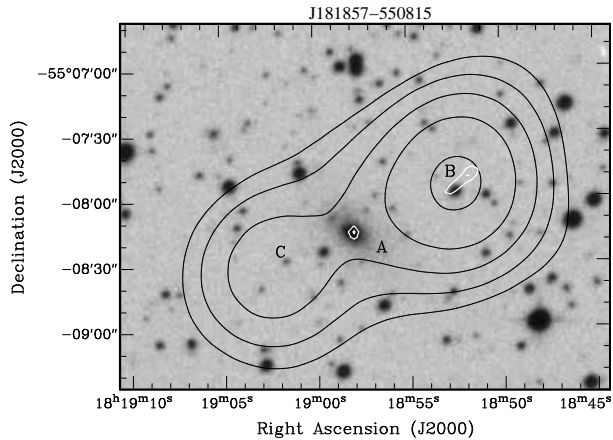, width=8cm}
 \caption{J181857$-$550815: SuperCOSMOS blue optical grayscale overlaid with black SUMSS radio contours at 20, 40, 80, 160 and 320 mJy/beam and white AT20G 4.8\,GHz radio contours at 30 and 40\,mJy/beam. The galaxy labeled A is J181857$-$550815. The objects labeled B and C are possible sources of the radio emission.}
 \label{fig:J181857-550815_optical}
\end{figure}

The 6dFGS spectrum of this galaxy (marked as object A in Figure  \ref{fig:J181857-550815_optical}) 
shows stellar absorption lines typical of early--type galaxies but no obvious optical emission lines. 
This galaxy lies between two SUMSS sources, as shown in Figure \ref{fig:J181857-550815_optical}. 
The AT20G source J181857$-$550815 is at the position of galaxy A near the centre of the image.  

The most likely interpretation is that the AGN in galaxy A produces the extended radio lobes 
which are seen at 843\,MHz. 
The radio observations at 5, 8 and 20GHz from the AT20G show an inverted spectrum 
with spectral index $\alpha_5^{20} = +0.40$ centered on J181857$-$550815, which may mean 
that, like J074618$-$570258, J181857$-$550815 is a recently ``restarted'' radio galaxy. 
The fact that the radio lobes are extended along the minor axis of the host 
galaxy is consistent with this scenario, and the 4.8\,GHz AT20G image shows a 
jet-like feature extending roughly 1\,arcmin from the nucleus. 

A second galaxy, marked as B in figure \ref{fig:J181857-550815_optical} lies close 
to the centroid of the north-western lobe and may also be responsible for 
some of the radio emission seen in the SUMSS image. 
The total SUMSS flux density is 627.7$\pm$21.2\,mJy 
and the separation of the two SUMSS components is 1.1\,arcmin, implying a largest 
linear size of at least 90\,kpc. 

\subsubsection{AT20G J220916$-$471000}
The AT20G source J220916$-$471000 is a well--studied nearby Seyfert 1 galaxy, NGC\,7213.   
The optical spectrum includes many strong emission lines which can be used as diagnostics of the 
physical conditions in the nucleus (Filippenko \& Halpern 1984).   Neutral hydrogen was also 
detected in this galaxy in the HIPASS survey (Doyle et al. 2005). 

The radio continuum emission from NGC\,7213 may be variable, as noted by Blank, Harnett and Jones (2005).


\subsubsection{AT20G J224506$-$433157}
The 6dFGS spectrum shows stellar absorption typical of an early-type galaxy, with no obvious emission lines. 
The optical counterpart of this source is only slightly extended. 

\section{Discussion}

\subsection{Source sizes} 
\begin{figure}
 \epsfig{file=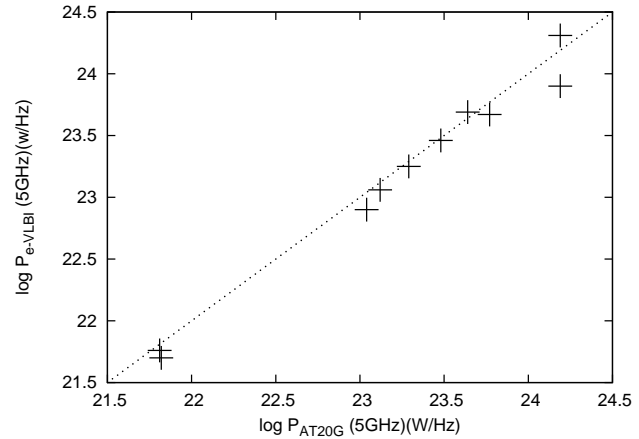, height=\linewidth,angle=-90}
 \caption{Relation between core and total radio power at 5GHz. The dashed line shows $P_c=P_t$. }
 \label{fig:radio_power}
\end{figure}

As noted earlier, all ten sources listed in Tables 1 and 2 were unresolved at $\sim$100\,mas resolution 
in our e-VLBI observations.  Figure \ref{fig:radio_power} shows that the radio emission detected by the AT20G survey on 
scales of 10--15\,arcsec or larger is dominated by a central compact component less than 
$\sim$0.1\,arcsec in size.  
The lack of any significant structure in the core on scales larger than 100\,pc 
supports the idea that we are looking at young unevolved radio sources. Even for a very slowly evolving 
source with hot spot expansion velocities of 0.1c, a linear size of \<100pc gives an age of \<3000 years.

For powerful radio galaxies, it is well known that core and total radio power are related 
(Fabiano et al. 1984), and Slee et al. (1994) found 
$P_c\propto P_t^{0.73\pm 0.05}$ at 5\,GHz for a sample of 140 galaxies ranging in 
radio power from $10^{20}$ to $10^{26}$\,W\,Hz$^{-1}$. 
Figure \ref{fig:radio_power} suggests that a similar relation applies for the 
nearby GPS radio galaxies in our sample.  


\subsection{Core spectral index}   
The distribution of the spectral indices $\alpha_5^8$ listed in Table \ref{table:observed_targets} 
is consistent with that of the medium--power sample studied by Slee et al. (1994). 
These authors attribute 
the inverted spectral indices of the galactic cores to a combination of synchrotron self 
absorption (SSA) and free-free absorption (FFA) for sources smaller than $\sim1$\,mas. 
For SSA to be responsible for the spectral turnover, the magnetic fields must be either 
much stronger than previously thought, or well below equipartition values, with the 
energy in relativistic electrons greatly exceeding that in magnetic fields. 
Orienti et al. (2008) find that equipartition does hold for high frequency peaking sources, 
requiring stronger magnetic fields than previously estimated by Slee et al. (2004). 
For sources larger than $\sim1$\,mas SSA is no longer viable. Higher--resolution 
observations of our GPS sample are needed to determine the angular size and 
structure of the central emission region.

\subsection{Variability}
GPS galaxies are the least variable class of compact radio sources (Rudnick and Jones, 1982).
GPS radio sources in general show a low incidence of variability 
($\sim 10\%$, O'Dea 1998 and references therein), and many of the high frequency peaking (HFP) sources are QSOs that 
show peaked spectrum only during outburst/flare events. This is supported by Torniainen 
et al. (2005) who find that nearly all quasar type GPS sources are variable both in spectral 
shape and radio power with only a small fraction being 'genuine' GPS sources.  
As our sample of sources contains only galaxies we might therefore expect very little variability 
to be present, but longer-term monitoring is needed to test this. At least two objects in the 
sample, J051103$-$255450 and J220916$-$471000, already show some evidence of variability, as noted in \S4.

\subsection{Extended low--frequency radio emission} 
At least two of the sources in our sample, J074618$-$570258 and J181857$-$550815, show extended low--frequency 
radio emission on scales of 100\,kpc or larger.  These may be ``restarted'' 
radio galaxies in which the current phase of nuclear activity has been caught at an early stage. 

It is particularly remarkable that our GPS sample, selected at 20\,GHz, includes the giant radio 
galaxy J074618$-$570258 which was first identified by Saripalli et al. (2005) on the basis of its 
extended low surface-brightness radio emission at 843\,MHz.  When the redshift coverage of the 
AT20G GPS sample is completed, it should allow a more detailed study of the duty cycle of 
activity in nearby radio galaxies.

\section{Conclusions and Future work}
We have presented 6cm e-VLBI observations of 10 low redshift, low radio power GPS galaxies 
selected from the AT20G survey. 
The angular resolution of the eVLBI observations was sufficient to
confirm the compact nature of the targets, but not high enough to
differentiate between edge-brightened and jet dominated GPS sources.
Such a differentiation is required to investigate the possibility of a
luminosity-morphology relationship in radio galaxy progenitors,
similar to the FR-I/FR-II relationship.  The eVLBI observations do
allow us to devise follow-up VLBI observations using the full LBA at a
higher observing frequency, to obtain higher angular resolution.  The
value of eVLBI observations in this context is that fast feedback can
be obtained regarding the detectability of the targets, allowing the
rapid selection of a sample for more detailed follow-up observations.

\section*{Acknowledgements}
The Australia Telescope Long Baseline Array is part of the Australia Telescope which is 
funded by the Commonwealth of Australia for operation as a National Facility managed by CSIRO.  
This research has made use of the NASA/IPAC Extragalactic Database (NED) which is operated by 
the Jet Propulsion Laboratory, California Institute of Technology, under contract with the 
National Aeronautics and Space Administration.

\section*{References}
Blank D.~L. et al. 2005, MNRAS, 356, 734 \\
Brosch N. et al., 2000, MNRAS, 313, 641 \\
Condon J.~J. et al., 1998, AJ, 115, 1693 \\
Deller A.~T. et al., 2007, PASP, 119, 318 \\
de Vaucouleurs, G., de Vaucouleurs A., Corwin H.G., Buta R.J., Paturel, G., Fouque P., Third Reference Catalogue of Bright Galaxies. \\
Doyle M.T. et al. 2005, MNRAS, 361, 34 \\
Fabiano G., Miller L., Trinchieri G., Longair M., Elvis M., 1984, ApJ, 277, 115\\
Fanaroff B.L., Riley J.M., 1974, MNRAS,167, 31P \\
Filippenko A.V., Halpern J.P., 1984, ApJ 285, 458 \\
Franx M., Illingworth G.D., 1988, ApJ 327, L55 \\ 
Griffith M.~R. et al., 1994, ApJS, 90, 179 \\
Hambly N.C. et al., 2001, MNRAS, 326, 1279  \\ 
Hancock P.J., 2009, AN, 330, 180 {\tt arXiv:0901.4592v1}\\
Kellerman K. 1966, AuJPh, 19, 195 \\
Jauncey D.~L et al. 2003, PASA, 20, 151 \\
Jones H. et al. 2004, MNRAS, 355, 747 \\
Malin D.~F., Carter D. ApJ, 274, 534 \\
Mauch T. et al. 2003, MNRAS, 342, 1117 \\
O'Dea C. 1998, PASP, 110, 493 \\
Orienti M. et al. 2008, A\&A, 487, 885\\
Rudnick L., Jones T.~W. 1982, ApJ, 255, 39 \\
Sadler E.~M et al. 2008, MNRAS, 385, 1656 \\
Saripalli L. et al. 2005, AJ, 130, 896 \\
Shepherd M. C. et al. 1994, BAAS, 26, 987 \\
Slee O.~B, et al. 1994, MNRAS, 269, 928 \\
Snellen I.~A.~G et al. 2000, MNRAS, 319, 445 \\
Stanghellini C., 2003, PASP, 20, 118 \\
Tingay S.~J et al. 2003, MNRAS, 346, 327 \\
Tingay S.~J., de Kool M. 2003, AJ, 126, 723 \\
Tzioumis A.~K, 1997, Vistas Astron., 41, 311 \\
Torniainen I. et al. 2005, A\&A, 435, 839 \\
Torniainen I. et al. 2007, A\&A, 469, 451 \\
Vermeulen R.~C et al. 2003, A\&A, 401, 113 \\
\label{lastpage}
\end{document}